\documentclass[a4paper,floatfix,superscriptaddress,prx,twocolumn,showpacs,notitlepage,longbibliography,nofootinbib]{revtex4-1}

\usepackage[colorlinks=true,citecolor=blue,urlcolor=blue]{hyperref}
\usepackage{graphicx}
\usepackage{dcolumn}
\usepackage{bm}
\usepackage[normalem]{ulem}

\usepackage[utf8]{inputenc}
\usepackage[T1]{fontenc}
\usepackage[sc,osf]{mathpazo}
\usepackage{amsmath, amsthm, amssymb,amsfonts,mathbbol,amstext}
\usepackage{graphicx}
\usepackage{dcolumn}
\usepackage{bm}
\usepackage{bbm}
\usepackage{hyperref}
\usepackage{mathtools}
\usepackage{comment}
\usepackage{color}
\usepackage{multirow}
\usepackage{diagbox}

\usepackage{epsfig, graphicx,psfrag,amsmath,amssymb,float}
\usepackage[T1]{fontenc}
\usepackage{amsmath}
\usepackage{amssymb}

\usepackage{amscd}
\usepackage{bm}
\usepackage{psfrag}
\usepackage{bbm} 
\usepackage{babel}
\usepackage{xcolor}

\newcommand{\be}{\begin{equation}}
\newcommand{\ee}{\end{equation}}
\newcommand{\bea}{\begin{eqnarray}}
\newcommand{\eea}{\end{eqnarray}}

\usepackage{braket}

\begin{document}

\title{Detecting quantum phase transitions in a frustrated spin chain via transfer learning of a quantum classifier algorithm}


\author{André J. Ferreira-Martins}
\affiliation{International Institute of Physics, Federal University of Rio Grande do Norte, 59078-970 Natal, Brazil}

\author{Leandro Silva}
\affiliation{International Institute of Physics, Federal University of Rio Grande do Norte, 59078-970 Natal, Brazil}
\affiliation{Instituto de Física de São Carlos, Universidade de São Paulo, CP 369, 13560-970 São Carlos, SP, Brazil}

\author{Alberto Palhares}
\affiliation{International Institute of Physics, Federal University of Rio Grande do Norte, 59078-970 Natal, Brazil}
\affiliation{Departamento de F\'isica Te\'orica e Experimental, Federal University of Rio Grande do Norte, 59078-970 Natal, Brazil}

\author{Rodrigo Pereira}
\affiliation{International Institute of Physics, Federal University of Rio Grande do Norte, 59078-970 Natal, Brazil}
\affiliation{Departamento de F\'isica Te\'orica e Experimental, Federal University of Rio Grande do Norte, 59078-970 Natal, Brazil}

\author{Diogo O. Soares-Pinto}
\affiliation{Instituto de Física de São Carlos, Universidade de São Paulo, CP 369, 13560-970 São Carlos, SP, Brazil}

\author{Rafael Chaves}
\affiliation{International Institute of Physics, Federal University of Rio Grande do Norte, 59078-970 Natal, Brazil}
\affiliation{School of Science and Technology, Federal University of Rio Grande do Norte, 59078-970 Natal, Brazil}

\author{Askery Canabarro}
\email{askery@gmail.com}
\affiliation{Instituto de Física de São Carlos, Universidade de São Paulo, CP 369, 13560-970 São Carlos, SP, Brazil}
\affiliation{Grupo de F\'isica da Mat\'eria Condensada, N\'ucleo de Ci\^encias Exatas - NCEx, Campus Arapiraca, Universidade Federal de Alagoas, 57309-005 Arapiraca-AL, Brazil
}

\date{\today}
\begin{abstract}

The classification of phases and the detection of phase transitions are central and challenging tasks in diverse fields. Within physics, it relies on the identification of order parameters and the analysis of singularities in the free energy and its derivatives. Here, we propose an alternative framework to identify quantum phase transitions. Using the axial next-nearest neighbor Ising (ANNNI) model as a benchmark, we show how machine learning can detect three phases (ferromagnetic, paramagnetic, and a cluster of the antiphase with the floating phase). Employing supervised learning, we demonstrate the feasibility of transfer learning. Specifically, a machine trained only with nearest-neighbor interactions can learn to identify a new type of phase occurring when next-nearest-neighbor interactions are introduced. We also compare the performance of common classical machine learning methods with a version of the quantum nearest neighbors (QNN) algorithm.  


\end{abstract}
\maketitle


\section{\label{sec:intro}INTRODUCTION}


Machine Learning (ML) has proven its efficiency and success in many scientific as well as business sectors \cite{aaron,highbias,altman,tpot,kelleher,charnock,casarini,carrillo, torlai2018neural,PhysRevLett.124.010508,carleo2017solving,torlai2016learning,ghiringhelli2015big,Carrasquilla,broecker2017machine,PhysRevX.7.031038,deng2017machine,huembeli2018identifying,goes2021}. In essence, we can teach computers to see patterns by progressively exposing them to quality inputs, which is crucial for data-driven solutions given the gigantic and ever-increasing amount of raw data. Within any branch of ML, substantial improvements in state-of-the-art solutions are strongly related to algorithmic and hardware advances. And, although we still need to be careful about setting long-term expectations, recent breakthroughs in the current noisy intermediate-scale quantum (NISQ) era \cite{qsupremacy,nisq,qf,qaoa} put quantum computing among the most promising directions towards significant progress in machine learning. Within this context, there has been a number of different approaches seeking for quantum advantages in machine learning ranging from the quantum analog of neural networks \cite{schuld2014quest}, routines for supervised or unsupervised learning \cite{lloyd2013quantum,wiebe2015quantum}, quantum reinforcement learning \cite{dong2008quantum} and quantum pattern recognition \cite{horn2001algorithm} (see references \cite{schuld2015introduction,biamonte2017quantum,carrasquilla2021neural} for a detailed review). 

A field where machine learning has been particularly successful is that of quantum matter and quantum information. Classical machine learning techniques were used to perform quantum state tomography \cite{torlai2016learning}, to approximate the ground state of many Hamiltonians of interest \cite{carleo2017solving}, for the description of causal networks \cite{krivachy2020neural,d2023machine,polino2023experimental} and finding violations of Bell inequalities \cite{deng2017machine,canabarro2019machine}, among many other applications. Importantly, such classical methods have also been proven capable of tackling a central topic in many-body physics, that of classifying phase transitions, a thorny goal especially due to the exponential increase of Hilbert space describing quantum systems. Apart from simple transitions, witnessed by non-analyticities in order parameters, more general quantum phase transitions require large lattice sizes, a costly computational task for which a variety of classical machine learning techniques provide alternative and reliable approaches \cite{broecker2017machine,Carrasquilla,canaPRB,PhysRevX.7.031038}. It seems thus natural to consider whether quantum machine learning can also identify phase transitions. Indeed, machine learning based on hybrid quantum-classical variational circuits has been shown to detect phase transitions in the simple Hamiltonians, such as the transverse field Ising and XXZ models \cite{uvarov2020machine,lazzarin2021multi,PhysRevB.66.064413,monaco}. Our approach distinguishes itself significantly from others, primarily through the implementation of transfer learning using a quantum classifier algorithm. This algorithm is exclusively trained within a specific segment of the phase diagram while testing on the rest of the phase diagram. We also explore optimized data preprocessing for compatibility with real quantum hardware. This demonstrates the effectiveness of our technique, as discussed in detail herein. 


Our aim in this paper is to show that the Quantum Nearest Neighbours (QNN) algorithm \cite{schuld} also provides a well-founded tool for classifying quantum phase transitions. Moving beyond the  models previously considered, we benchmark the QNN performance by employing the axial next-nearest neighbor Ising (ANNNI) model \cite{Selke,Suzuki} used, for instance, to investigate the magnetic order in quasi-one-dimensional spin ladder materials \cite{Wen}, quantum melting of crystalline order in Rydberg atom systems \cite{Sela2011}, interactions between Majorana edge modes in arrays of Kitaev chains \cite{Hassler,Milsted}, and quench dynamics and dynamical phase transitions \cite{Karrasch,Robertson2023,Haldar2023}. The ANNNI is the simplest model combining the effects of quantum fluctuations and frustrated exchange interactions, a combination from which a rich ground state phase diagram arises \cite{Villain,Allen,Rieger, Guimaraes,Beccaria,Nagy}. It thus provides an interesting challenge to the QNN algorithm capabilities.

Importantly, even though the input data to the quantum algorithm is considerably small, the full $198$ raw pairwise correlation functions between the spins for a lattice with 12 sites, i.e., an input array of $198$ real features. 
And even if each of these variables is mapped to just one bit, we would still require a large amount of qubits. As better detailed ahead, to be implemented, the QNN algorithm that we use requires $2n + 2$ qubits for a $n$-sized input vector. In order to make the computational simulation as well as its implementation in a real quantum computer feasible, we first proceed with a pre-processing of the data, consisting of a feature selection followed by a discretization and a final one-hot encoding step. With that, we reduce to $4$ the number of features that in turn require $10$ qubits to be analyzed by the quantum algorithm.

As we show, even after significantly reducing the input data, to make it compatible with quantum computational requirements, the QNN algorithm still allows for a successful learning of phase transitions. More precisely, we demonstrate the transfer learning in the ANNNI model, as by training the machine with nearest-neighbour interactions only, it also accurately predicts the phase transitions happening at regions including next-nearest-neighbor interactions. Interestingly, the QNN performs better than its classical counterpart, called K-Nearest Neighbors (KNN), when exposed to the same input data, thus providing a proof-of-principle example of a possible quantum advantage in accuracy.

The paper is organized as follows. In Sec. II we
describe the ANNNI model. In Sec. III we
provide a succinct but comprehensive overview of classification problems in machine learning, also describing the KNN and QNN algorithms. In Sec. IV we detail the data pre-processing required to make the problem amenable to be implemented in near-term quantum devices. In Sec. V we present our results regarding the learning of phase transitions in the ANNNI model. In Sec. VI we discuss our results and point out interesting directions for future research. Finally, in the Appendix
we provide technical details about some of the classical machine learning techniques we have employed in the pre-processing of the data.

%

\section{\label{sec:model}The ANNNI Model}

With the goal of analyzing the use of a quantum classifier algorithm to witness phase transitions in a quantum many-body system, we chose the axial next nearest-neighbor Ising (ANNNI) model. The reason stems from the fact that this model displays a non-trivial and rich phase diagram.  As it happens, ANNNI is the simplest model combining quantum fluctuations and competing frustrated exchange interactions. The first is induced by the presence of a transverse field while the latter is due to the fact that even though the interaction is ferromagnetic for nearest neighbors, it becomes antiferromagnetic for next-nearest neighbors.

The Hamiltonian for the ANNNI model is given by \cite{Selke,Suzuki}
\be
H=-J\sum_{j=1}^N\left(\sigma_j^z\sigma_{j+1}^z-\kappa \sigma_j^z\sigma_{j+2}^z+g \sigma_j^x\right),\label{ANNNI}
\ee
where $\sigma^{\alpha}_j$ ($\alpha=x,y,z$), are Pauli matrices acting on spin-$1/2$ degrees of freedom at site $j$  of a one-dimensional lattice with $N$ sites and periodic boundary conditions. The parameter $J>0$ is a coupling constant that sets the energy scale of the problem (we set $J=1$) and is associated with the nearest-neighbor ferromagnetic exchange interaction. The dimensionless coupling constants $\kappa$ and $g$ are related to the next-nearest-neighbor interaction and the transverse magnetic field, respectively. 

The groundstate phase diagram of the ANNNI model is well understood and known to contain four phases separated by three quantum phase transitions: ferromagnetic, antiphase, paramagnetic, and floating phase. In a nutshell,  the ferromagnetic phase is characterized by a uniform spontaneous magnetization, with one of the ground states given by $\uparrow\uparrow\uparrow\uparrow\uparrow\uparrow\uparrow\uparrow$. In turn, the antiphase breaks the lattice translational symmetry and has long-range order with a  four-site periodicity of the form  $\uparrow\uparrow\downarrow\downarrow\uparrow\uparrow\downarrow\downarrow$. Distinctively, the paramagnetic phase is disordered and has a unique ground state with spins pointing predominantly along the field direction. Finally, the floating phase is gapless with correlation functions decaying as a power-law for large distances, in contrast with the other phases that have a finite energy gap and exponentially decaying correlations.

For $\kappa=0$, the transverse field Ising model is reproduced, exactly solvable via the mapping to non-interacting spinless fermions. Along the $\kappa=0$ line, a second-order phase transition occurs at $g=1$, separating the ferromagnetic phase at $g<1$ from the paramagnetic phase at $g>1$. In particular, exactly at the critical point $g=1$, the energy gap vanishes. 

For $g=0$, there is a  transition between the ferromagnetic phase at small $\kappa$ and the antiphase at large $\kappa$ occurring at $\kappa=1/2$. Notice that with $g=0$ the model becomes classical, since all operators in the Hamiltonian commute with each other. At this classical transition point, any configuration that does not have three neighboring spins pointing in the same direction is a ground state, showing that the degenerescence of the ground state increases exponentially with the system size.

For $g\neq 0$ and $\kappa\neq0$, the critical lines have to be determined numerically since the model is not integrable any longer. For $0\leq\kappa\leq1/2$, the Ising transition between paramagnetic and the ferromagnetic phases extends from the $g=1$, $\kappa=0$ until the degenerate point $g=0$, $\kappa=1/2$, a multicritical point at which several transition lines coincide. There are two other transition lines that start at the multicritical point and extend to the high-frustration regime $\kappa>1/2$. For fixed $g>0$ and increasing $\kappa>1/2$, we first encounter a Berezinsky-Kosterlitz-Thouless (BKT) transition from the paramagnetic phase to the floating phase. Detecting the BKT transition is challenging  because the correlation length diverges exponentially at the critical point. As we increase $\kappa$ further, there is a commensurate-incommensurate (CIC) transition from the floating phase to the antiphase. Numerical density matrix renormalization group results for long spin chains \cite{Beccaria} show that the floating phase occupies a rather narrow region in the phase diagram, which makes it hard to discern the BKT  from the CIC transition for small system sizes. 

Using perturbation theory in the regime $\kappa<1/2$ \cite{Suzuki} or by fitting numerical results  \cite{Beccaria}) in the regime $\kappa>1/2$, one can obtain approximate expressions for the transition lines. For instance, the critical value of $g$ for the Ising transition  for $0\leq \kappa\leq 1/2$ is approximately given by \cite{Suzuki}
\be
\label{trans1}
g_{\textrm{I}}(\kappa)\approx \frac{1-\kappa}{\kappa}\left(1-\sqrt{\frac{1-3\kappa+4\kappa^2}{1-\kappa}}\right).
\ee
In turn, the critical value of $g$ for the BKT transitions for $1/2<\kappa\lesssim 3/2$ is approximated by \cite{Beccaria}
 \bea
 \label{transBKT}
 g_{\textrm{BKT}}(\kappa)&\approx& 1.05\sqrt{ (\kappa-0.5 ) (\kappa-0.1)}.
 \eea
We use these approximations to make benchmark comparisons to our heuristic results.

\subsection{\label{sec:dataset}Our dataset}
As will be discussed in more detail throughout the paper, we use the pairwise correlations among all spins in the lattice as the training data for the machine learning algorithms. Given $N$ spins, 
we have a total of $3
\times C_2 = \binom{N}{2} $ observables for up to (second) nearest neighbors, hence the combination by (2). Thus, the features are given by $\left\{ \langle \sigma^{x}_{i}\sigma^{x}_{j} \rangle,\langle \sigma^{y}_{i}\sigma^{y}_{j } \rangle, \langle \sigma^{z}_{i}\sigma^{z}_{j} \rangle \right\}$ with, $j>i$ and $i=[1,N-1]$ where $N$ is the number of spins in the lattice and $\langle \sigma^{x}_{i}\sigma^{x}_{j} \rangle =\langle\lambda_{0} \vert \sigma_{i}^{x}\sigma_{j}^{x} \vert \lambda_{0}\rangle$ is the expectation value of the spin correlation for the Hamiltonian ground state $\vert \lambda_{0}\rangle$ (and similarly for the other expectation values). 
In our case, we take N = 12, a manageable size for both computational and analytical evaluation of the ground state of the ANNNI Hamiltonian. This allows us to efficiently compute a set of 198 pairwise expectation values, which will serve as the (raw) input features for the machine learning algorithm.

It is worth pointing out that, even if one only has access to short chains, the Ising transition can still be captured correctly \cite{Guimaraes}. However, detecting the BKT transitions using standard approaches requires computing observables for significantly longer chains \cite{Beccaria}. Notwithstanding, as we will see below, even though our data regards a quite short chain $N=12$, the machine learning algorithms, both classical and quantum, will be able to identify not only the Ising but also the antiphase and the paramagnetic phases, lumping the BKT and CIC transitions together. 


\section{The Quantum nearest neighbors algorithm}
\label{sec:QNN}


The quantum nearest neighbors (QNN) \cite{schuld} is a quantum classification algorithm that employs the Hamming distance as a distance criterion to compare the training dataset and unclassified observations. Schematically, it consists of three major steps:
\begin{itemize}
    \item First, create a superposition of the training dataset and the input observation;
    \item Encode the Hamming distance between the input observation and each example in the training set into the amplitude of each state in the superposition;
    \item Measure the class-qudit retrieving the predicted class with the highest probability.
\end{itemize}
  

Before the actual quantum algorithm starts, an important classical pre-processing step (whose reason will become clear in what follows) must be performed: the features in the training dataset are represented as bit vectors, so that the feature space becomes $\mathcal{X} = \{0, 1\}^{\otimes n}$. This is achieved via the procedure known as one-hot encoding, which produces the so-called dummy variables \cite{dummy}. Naturally, such a representation will be discrete (binary, in fact), so that if any original feature is continuous (or even categorical with more than 2 levels), a prior step of discretization is necessary. Notice that the number of binary features after the encoding may be different from the number of original features, although here we represented both by the same number of bits $n$. There are several ways to perform this binarization process. However, whatever method is chosen, it is important that the essence of the data topology is maintained --- that is, points that are close on the original feature space must remain close on the binarized feature space. In Sec. \ref{sec:methods_qnn}  we detail the specifics of the particular procedure we applied to our problem.

Once the training dataset features are binarized, their representation as quantum states is immediate via the basis encoding \cite{encode}, which accounts for a direct mapping of binary features to the quantum computational-basis states: $0 \mapsto \ket{0}$ and $1 \mapsto \ket{1}$.
After these two steps, each training set data point is mapped to the quantum state $\ket{x_1^p \cdots x_n^p} \equiv \ket{\bm{x}^p}$, $x_k^p \in \{0,1\} $, $p = 1, \cdots, N$, where $N$ is the number of points in the training set.
In parallel, in a separate quantum register, we encode the class $y^p \in \{0, \cdots, d-1\}$, and construct, for each observation $p$, the state
\begin{equation}
 \ket{x_1^p \cdots x_n^p, y^p} \equiv \ket{\bm{x}^p, y^p} 
 \ .
\end{equation}

If we are dealing with binary classification (which is the case in this work), the respective class is also straightforwardly encoded in a single qubit, as $0 \mapsto \ket{0}$ and $1 \mapsto \ket{1}$. If we have a multiclass problem, qudits are necessary, or one could use more than one qubit to encode integers corresponding to the class (for instance, $\ket{5} = \ket{101}$). In this case, $\lceil \log_2d \rceil$ qubits are necessary to encode $d$ classes.

Once we have the state corresponding to each one of the training states 
$\ket{\bm{x}^p, y^p}$, we construct a training set superposition of all datapoints, given by
\begin{equation}
   \ket{T} = \frac{1}{\sqrt{N}} \sum_{p=1}^N \ket{\bm{x}^p, y^p}  \ .
\label{eq:train_sup}
\end{equation}

Naturally, with $n$ qubits, one can construct a superposition of $2^n$ states, representing all possible binarized feature vectors of $n$ features. However, it is possible (and most likely) that in a given training dataset not all possible binary feature vectors will be present. Indeed, in the binarization process, it is likely that multiple observations that are different in the original input space are mapped to the same binary vector so that the transformed training dataset actually has a number of observations quite smaller than the original number of observations (although here we represent both as $N$). This leads to important details in the implementation of the algorithm in a practical problem, as it will be detailed in Sec. \ref{sec:methods_qnn}. Further, notice that in the case in which $N < 2^n$, the superposition in Eq. \eqref{eq:train_sup} will have to be prepared with an arbitrary state preparation routine, which is known to be costly \cite{state_prep}. However, in quantum computing software development kits (SDK) (such as Qiskit \cite{qiskit}, which is the one we employ in this work), such a procedure is already implemented and ready to use as a self-contained routine.

The next step is to perform the same classical binarization process with the unclassified input vector $\bm{x_{\text{in}}}$ (the one we wish to classify) and map it to the state $ \ket{x_{\text{in},1} \cdots x_{\text{in},n}} \equiv \ket{ \bm{x_{\text{in}}}}$, $x_{\text{in},k} \in \{0,1\}$. Keep this as the first register of the quantum state. Finally, add an ancilla register $\ket{0}$ as the last register. Such a construction yields an initial state given by
\begin{equation}
    \ket{\psi_0} = \frac{1}{\sqrt{N}} \sum_{p=1}^N \ket{\bm{x_{\text{in}}};\bm{x}^p, y^p ; 0} \ , 
\label{eq:psi0}
\end{equation}
which is made up of three registers (or, in fact, blocks of registers): the first containing the input state $\ket{\bm{x_{\text{in}}}}$, which consists of $n$ qubits; the second containing the superposition $\ket{T}$ (which is the tensor product of the feature vectors $\ket{\bm{x}^p}$ and the class vectors $\ket{y^p}$), thus consisting of $n+1$ qubits, and given that we have a binary classification problem, the third contains an ancilla qubit initialized as $\ket{0}$. Therefore, the number of qubits necessary for the algorithm is precisely $2n + 2$.

Once the initial state is prepared, we put the ancilla into a superposition, by applying a Hadamard gate to the last register, i.e., $H = 1 \otimes 1 \otimes 1 \otimes H$, such that
\begin{equation}
   \ket{\psi_1} = H \ket{\psi_0} = \frac{1}{\sqrt{N}} \sum_{p=1}^N \ket{\bm{x_{\text{in}}};\bm{x}^p, y^p} \otimes \frac{1}{\sqrt{2}} (\ket{0}+ \ket{1}) \ .
\label{eq:psi1}
\end{equation}

In the next step, we want the Hamming distance components $d_k^i$ between each qubit of the first (input) and second (training) register to replace the qubits in the second register, such that
\begin{equation}
d_k^i = 
\begin{cases}
0 , & \text{ if } \ket{x_k^p} = \ket{x_{\text{in},k}} \\ 
1 , & \text{ else. } 
\end{cases} \ .
\label{eq:hamming}
\end{equation}
	
This is achieved by simply applying a $\mathrm{cNOT}(x_{\text{in},k},x_k^p)$-gate, which overwrites the entry $x_k^p$ in the second register with $0$ if $x_k^p=x_{\text{in},k}$, otherwise with $1$:
\begin{equation}
\begin{cases}
\mathrm{cNOT} \ket{00} = \ket{00} \ ; & \  \mathrm{cNOT} \ket{01} = \ket{01}\\ 
\mathrm{cNOT} \ket{11} = \ket{10} \ ; & \ \mathrm{cNOT} \ket{10} = \ket{11}
\end{cases} \ . 
\end{equation}

Thus, after this step, the state is then 
\begin{equation}
\begin{aligned}
\ket{\psi_2} &= \bigotimes_{k=1}^n \mathrm{cNOT}(x_k, v^p_k) \; \ket{\psi_1}
\\
&= \frac{1}{\sqrt{N}}\sum_{p=1}^N \ket{\bm{x_{\text{in}}};\bm{d}^p, y^p} \otimes \frac{1}{\sqrt{2}}(\ket{0} + \ket{1}) \ ,
\label{eq:psi2}
\end{aligned}
\end{equation}
where the Hamming distance components $\ket{d_1^p \cdots d_n^p} \equiv \ket{\bm{d}^p}$, $d_k^p \in \{0,1\}$, $p = 1, \cdots, N$ are now in the second register.


In the third step, we apply the unitary operator
\begin{equation}
U = e^{ -i\frac{\pi }{2n} \mathcal{O}}  \ ; \mathcal{O} = 1 \otimes \sum_{k=1}^n \left(\frac{1-\sigma_z}{2}\right)_{d_k}  \otimes 1 \otimes \sigma_z\; \ .
\label{eq:U_op}
\end{equation}
This sums the Hamming distance components $\{d^p_k\}$ (thus yielding the actual Hamming distance) between $\ket{\bm{x}^p}$ and $\ket{\bm{x_{\text{in}}}}$, $d_H(\bm{x_{\text{in}}},\bm{x}^p) \equiv d_H$, into the phase of the $p$\textsuperscript{th} state of the superposition. Notice that a relative phase is added, conditioned on the ancilla state. After this step, the state becomes
\begin{widetext}
\begin{equation}
\ket{\psi_3} = U \ket{\psi_2} = \frac{1}{\sqrt{2N}} \sum_{p=1}^N  \left ( e^{ -i\frac{\pi }{2n} d_H}  \ket{\bm{x_{\text{in}}};\bm{d}^p, y^p; 0} + e^{i\frac{\pi }{2n} d_H} \ket{\bm{x_{\text{in}}};\bm{d}^p, y^p; 1} \right ) \ .
\label{eq:psi3}
\end{equation}
\end{widetext}

Now we apply another Hadamard to the ancilla. This will generate alternating-sign exponentials associated with each ancilla state, which are easily aggregated into a sine and cosine. The resulting state can be expressed as 
\begin{widetext}
\begin{equation}
\ket{\psi_4} = H \ket{\psi_3} = \frac{1}{\sqrt{N}} \sum_{p=1}^N  \left ( \cos\left (\frac{\pi d_H}{2n}\right) \ket{\bm{x_{\text{in}}};\bm{d}^p, y^p;0} + \sin\left (\frac{\pi d_H}{2n}\right) \ket{\bm{x_{\text{in}}};\bm{d}^p, y^p;1} \right ) \ .
\label{eq:psi4}
\end{equation}
\end{widetext}

Notice that $0 \leq d_H  \leq n \Rightarrow 0 \leq \frac{\pi d_H}{2n} \leq \frac{\pi}{2}$. Therefore,
\begin{itemize}
    \item For large $d_H$, $\cos\left (\frac{\pi d_H}{2n}\right) \rightarrow 0$ and $\sin\left (\frac{\pi d_H}{2n}\right) \rightarrow 1$, so that we have higher probability of measuring $\ket{1}$ in the ancilla qubit;
    \item For small $d_H$, $\cos\left (\frac{\pi d_H}{2n}\right) \rightarrow 1$ and $\sin\left (\frac{\pi d_H}{2n}\right) \rightarrow 0$, so that we have higher probability of measuring $\ket{0}$.
\end{itemize}

That is, if the input is far away from most training observations, we have a higher probability of measuring the ancilla in the state $\ket{1}$; and if the input is close to many observations, the ancilla is more likely to be measured in $\ket{0}$. Thus, intuitively, since our criterion for classification is to consider the closest observations, by measuring the ancilla in $\ket{0}$, the amplitudes of close observations will be large, whilst the opposite is true for distant observations. The importance of this fact becomes clear if we rewrite $\ket{\psi_4}$, to show that the different classes appear weighted by their member's distance to the input, such that
 \begin{widetext}
 \begin{equation}
\ket{\psi_4} = \frac{1}{\sqrt{N}} \sum_{y=0}^{d-1} \ket{y} \otimes \sum \limits_{l\in y} \left ( \mathrm{cos}\left( \frac{\pi d_H}{2n}\right)  \ket{\bm{x_{\text{in}}};\bm{d}^l; 0} +  \mathrm{sin}\left( \frac{\pi d_H}{2n}\right)  \ket{\bm{x_{\text{in}}};\bm{d}^l; 1} \right )  \ , 
\label{eq:psi4_rewritten}
\end{equation}
 \end{widetext}
where $l$  runs over all training vectors classified with the label $y$. Written like this, it becomes clear that, if the ancilla is measured to be in $\ket{0}$, the amplitudes of close observations will be large, which implies that the probability of measuring the respective class qubit of these observations will also be large. And, as we discuss next, this is how the final classification is performed. 

As the final step, the ancilla of the state $\ket{\psi_4}$ is measured on the computational basis. According to Eq. \eqref{eq:psi4}, it is easy to see that the probability of measuring $\ket{0}$ is
\begin{equation}
P(\ket{0}_a) = \left | \braket{0|\psi_4} \right |^2 =  \frac{1}{N} \sum_{p=1}^N  \cos^2\left (\frac{\pi d_H}{2n} \right) \ .
\label{eq:p_0}
\end{equation}    

The conditional probability to measure a certain class $y \in \{1,...,d\}$, given that we previously measured the ancilla in $\ket{0}$ (and, therefore, the state collapsed to $\ket{\tilde{\psi}_4} = \braket{0|\psi_4}\ket{0}$) is, in terms of the joint probability,
\begin{equation}
\begin{aligned}
P( y \mid \ket{0}_a ) &= P(y)P(\ket{0}_a)
\\
&= \left |  \braket{y|\tilde{\psi}_4} \right |^2
\\
&= \frac{1}{N} \sum \limits_{l\in y} \cos^2\left (\frac{\pi d_H}{2n} \right) \ ,
\label{eq:p_1}
\end{aligned}
\end{equation}    
which is easily verifiable using Eq. \eqref{eq:psi4_rewritten}. Indeed, Eq. \eqref{eq:p_1} implies that
\begin{equation}
P(y) =  \frac{1}{ P(\ket{0}_a)}\frac{1}{N} \sum \limits_{l\in y} \cos^2\left (\frac{\pi d_H}{2n} \right) \ .
\label{eq:p_final}
\end{equation}   
Thus, the class measured with the highest probability is that whose members are the closest to the input vector, provided that $P(c)$ is only computed after the ancilla is measured in $\ket{0}$, which is precisely why the amplitudes associated to the closest neighbors are considered. Notice that if the measurement returns $\ket{1}$, this run of the algorithm is not taken into account.

In Fig. \ref{fig:qc} the full quantum circuit is illustrated for a particular dataset, as detailed in Appendix \ref{sec:qc}. The algorithm uses $\mathcal{O}(Nn)$ \cite{schuld} gates, which is completely due to the construction of the training data superposition (described by Eq. \ref{eq:train_sup}), which depends on the number of training samples, thus yielding a $\mathcal{O}(Nn)$ complexity \cite{schuld, associative_memory}, which close to the classical KNN algorithm complexity, in which we have to compute the distance between the test observation and all other $N$ training points. However, if one can find a procedure to prepare the training data superposition in a manner independent of the number of samples (perhaps by reading quantum data \cite{quantum_data1, quantum_data2, quantum_data3, quantum_data4}), the QNN algorithms would run in $\mathcal{O}(n)$, offering a potentially large advantage over the classical KNN, for which it seems unlikely to exist an algorithm which is independent of the number of training samples --- that's a quite remarkable advantage achieved by the exploitation of the superposition in a quantum algorithm.


We highlight that, in contrast to the classical KNN, the QNN algorithm does not depend on the hyperparameter $k$. In fact, a superposition of all members of each class is taken into consideration for the classification. This is equivalent to considering all neighbors (that is, $k=N$), which in the classical algorithm is associated with a high bias, since, if the dataset is imbalanced with respect to the target, the majority class will always be predicted. In the quantum algorithm, however, this is not the case: even if the dataset is imbalanced, the input observation will be assigned to the class which is the closest to it, since, as it is clear in Eq. \eqref{eq:p_final}, the distance of the input to the members of the class explicitly affects the probability.

As a final remark, notice that the probability distribution in Eq. \eqref{eq:p_final} is precisely what is recovered from multiple runs of the algorithm on an actual hardware (or a simulator thereof). The final prediction, as a class, is therefore recovered by choosing the class with the largest observed probability. However, as explained in Sec. \ref{sec:methods_qnn} and illustrated in Fig. \eqref{fig:fig3}, in this work we directly use the class probability, i.e., Eq. \eqref{eq:p_final} itself. Fortunately, in contrast to many classical learning models, outputting class probabilities is the most natural choice for the QNN algorithm.

We remark that the Python/Qiskit implementation of the algorithm described above, as well as all the data used in this paper and the corresponding results, are available in an online repository \cite{repo_qml}.



\section{\label{sec:methods_qnn} Data pre-processing}

As described in Sec. \ref{sec:QNN}, the classical data loaded into the quantum registers, via the basis encoding strategy, must be in a binary representation. On the other hand, as discussed in Sec. \ref{sec:dataset}, the dataset under study consists of 198
continuous features: the pairwise correlations among all spins in a lattice with 12 sites considering boundary conditions and the symmetry it implies.
Thus, in order to represent each observation as a binary vector, we must first discretize the continuous features, so that the discrete levels may then be binarized.

Before proceeding with these procedures, however, an important remark is due. As discussed above, the QNN algorithm uses $2n+2$ qubits, where $n$ is the number of binarized features. Indeed, this implies that if one wants to simulate the circuit, or even execute it on NISQ hardware, $n$ must be chosen accordingly, to make the execution/simulation feasible.

As will be detailed below, we employed an efficient discretization and binarization procedure that maps each original continuous feature to only two bits. However, given that we start with 198 features, this would imply $n=396$, thus requiring a circuit consisting of $794$ qubits, which is way beyond the limit for a classical simulation (recall that the algorithm includes entanglement) as well as current quantum hardware capabilities, both in terms of number of qubits as well as circuit depth. And this is a quite simple discretization one can think of: if one produces more bins per feature, which would be desirable, the resulting number of binary features (and qubits, consequently) would further increase, making the simulation/execution yet more intractable. 

Therefore, in order to fit the dataset into current capabilities, we employ a series of pre-processing steps to the original raw features, which starts with a dimensionality reduction procedure implemented via a feature selection routine, in order to pick from the 198 original features, the ones that contribute the most to the classification, with the hope that they are enough to produce a good classifier. The procedure we use for picking the most important features is based on the Random Forest algorithm \cite{breiman} --- in particular, a modification thereof, known as Extremely Randomized Trees \cite{extra_trees}. It consists of the calculation of a "feature importance coefficient" which is also known as "mean decrease impurity" or "gini importance" \cite{feature_imp}. This coefficient is calculated as the total decrease in each node impurity, averaged over all trees in the forest. The mean is weighted by the probability that the respective node is reached, which is often estimated by the proportion of observations reaching the node. 
A detailed account of this algorithm can be found in the Appendix \ref{sec:rf}. 

\begin{figure}[!t]
\begin{center}
\includegraphics*[width=\linewidth]{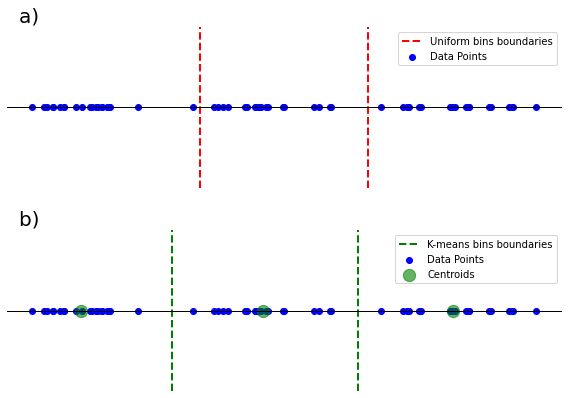}
\end{center}
\caption{\textbf{k-bins discretizer using uniform and k-means ($k=3$) binning strategies.} 
\textbf{a)} k-bins discretizer with bins uniformly defined. The vertical red lines represent the bins' limits. Notice that bin widths are uniform in the feature space, but the clustering structure is not respected. 
\textbf{a)} The green points represent the clusters centroids, and the vertical green lines, the bins' limits. Notice how non-uniform bins are created, but the clustering structure is respected. 
}
\label{fig:kbin_kmeans}
\end{figure}

Having in mind the discussion about the current capabilities of simulation and hardware, we have selected only the 4 most important features that correspond to the following two-body correlation terms. Physically, we expect that the most important features are the correlation functions  $\langle \sigma_i^z\sigma_j^z\rangle$ at the largest available distances. The reason is that this correlation detects long-range order associated with the spontaneous breaking of the $\mathbb Z_2$ symmetry $\sigma_j^z\mapsto -\sigma_j^z$ of the Hamiltonian. In the paramagnetic phase, $\langle \sigma_i^z\sigma_j^z\rangle$ decays exponentially to zero for $|i-j|$ larger than the correlation length, while it approaches a nonzero value in the ferromagnetic phase and oscillates with a four-site periodicity in the antiphase. By contrast, the correlation function $\langle \sigma_i^x\sigma_j^x\rangle$ is nonzero for $g\neq0$ in all phases because the transverse magnetic field induces a finite expectation value for $\sigma_j^x$.


We then proceed to the discretization of the features, using a procedure based on the $k$-means clustering algorithm \cite{kmeans}. 
More specifically, we use a k-bins discretizer, implemented in scikit-learn \cite{scikit}, which divides continuous data into $k$ intervals or "bins". Essentially, we first cluster observations that are similar to the feature being discretized and use the clusters centroids as centers of the bins, that is, values in each bin will have the same nearest centroid, as determined by the 1-dimensional $k$-means algorithm. See Fig. \ref{fig:kbin_kmeans} for an illustration.
For each feature, we created 3 bins. At this point, our dataset is characterized by 12 discrete values, 3 for each one of the \textcolor{red}{4} features selected by the feature importance procedure.

\begin{figure}[!t]
\begin{center}
\includegraphics*[scale=0.35]{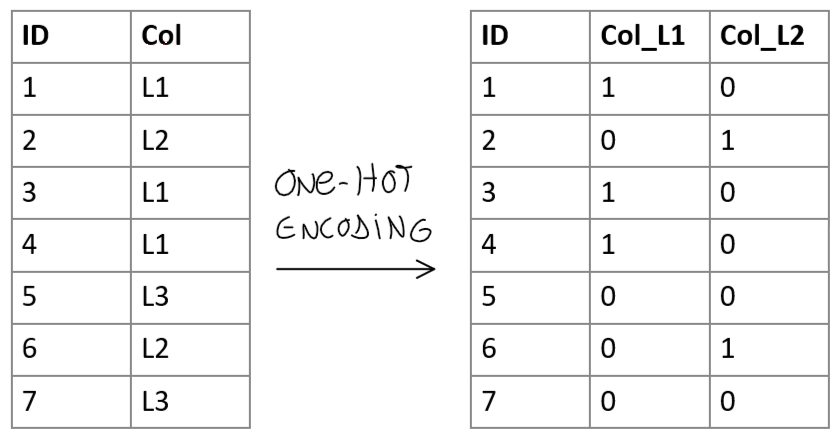}
\end{center}
\caption{\textbf{One-hot encoding procedure.} From a column with $l=3$ categorical (discrete) levels, we construct $l-1=2$ independent binary columns, which are the binary features. Notice that only $l-1$ binary columns are necessary because it is possible to represent one of the levels as "00" (in this example, "L3").}
\label{fig:one-hot}
\end{figure}

After discretization, the features are binarized using the one-hot encoding procedure, which consists of creating $l-1$ independent binary features for each column with $l$ categorical levels, as illustrated in Fig. \ref{fig:one-hot}. In our case, since $l=3$ for each discretized feature, we create $l-1 = 2$ new binary features each, which then results in $n=8$ independent binary features. This is the final dimensionality we work with.

Notice that, with 8 binary features, we will need 18 qubits for the circuit execution, which is a reasonable number for the circuit simulation or execution --- and, most importantly, it is enough to yield good results for the problem we analyze, as it will be shown. We could have chosen more than $n=4$ features or discretized the features in more bins, which could possibly have increased the performance quantum classifier.
In Sec. \ref{sec:disc} we further elaborate on this point.

Notice that after the feature selection, discretization and binarization pre-processing described above, some observations which were different in the original space of continuous features may be mapped to the same binary representation. This makes perfect sense, given that the different values of a given feature may fall in the same bin, which is given by a range of values. If this happens with all 4 features of two different observations, they will naturally be represented by the same 8-dimensional feature vector. This is why using a $k$-means binning strategy is a good idea (instead of the competing strategy "uniform", for example, in which all bins have identical widths, as depicted in Fig. \ref{fig:kbin_kmeans}b):
given that bins are clusters, this strategy groups together similar observations in the same bin, so that it makes sense if they are represented by the same binary feature vector.


After the pre-processing routine, our original training dataset, which had \textcolor{red}{1000} rows and 198 continuous-valued columns, was reduced to a dataset with 8 binary features and only 10 rows. 
We can see that as a way of reducing the dataset only to the most representative samples and better explanatory features, which, as we will show below, was enough to yield good results with the quantum classification algorithm. 


It is important to remark that the aforementioned feature importance and discretization processes were such that their respective parameters were determined only using the training data. That is, the exact same procedure was merely reproduced for all test datasets, but with parameters already defined in the training data.
Now, although this is the correct thing to be done to avoid data leakage, there is a potential problem, especially with the discretization process: given that the features range varies a lot from training to testing data, it is possible that the resulting bins for testing data features will be concentrated, that is, all observations will fall in the same bin of a given feature. Effectively, this implies that such test observations will be characterized by less than 8 features, which is a problem because the QNN algorithm assumes that the test (input) observation has the same number of features as all training observations. In order to fix this, we pad such input observations with zeros, to guarantee that all binarized testing observations will be 8-dimensional. In practice, different observations will be identified only by the actual features present, and the padding will have no effect in terms of the classification, given that it will be the same for all observations in a given test dataset, as we observed. Indeed, as the results show, such a procedure does not jeopardize the classifier's performance.

As already remarked, remember that QNN is a lazy algorithm, so each test (input) observation is classified at a time. This means that, in principle, we would have to simulate/execute a number of circuits equal to the number of test observations, to have their prediction. Given that we have 10 testing sets, one for each $k$ value. We consider $k = 0$ the training point.
each one with 1000 observations, the number of circuit simulations/executions would be quite large. However, the aforementioned fact that different training observations in the original feature space may be mapped to the same binary representation is of course also true for the testing data observations (although the exact number of unique binarized testing observations vary among the different test datasets). Given that, we implement a cache: whenever we see a new testing observation (in terms of its 8 features), we pass it through the quantum circuit, simulating/executing it, and store its prediction. If this observation is repeated (which, again, can happen given the nature of the pre-processing routine), we don't run the circuit again, but instead merely replicate the prediction in the cache. 
This allows us to have a prediction for each one of the observations in the testing datasets, without having to simulate/execute the quantum algorithm that many times. Indeed, this is very important for resource optimization, in terms of simulation time or hardware calls for execution. 

\section{Machine Learning the phase diagram of the ANNNI model with QNN}

\begin{figure}[t!]
\begin{center}
\includegraphics*[scale=0.6]{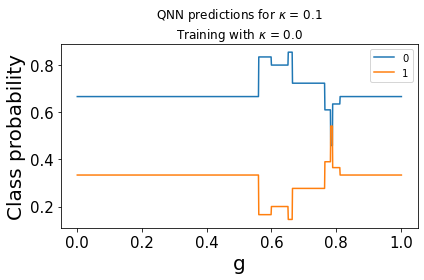}
\end{center}
\caption{Detecting the critical transverse magnetic field coupling parameter $g$ at which a phase transition occurs. The machine was trained at $\kappa=0$ and asked to predict where the transition happens at $\kappa=0.1$, by considering where the machine is most uncertain, that is, when the probabilities are closest $p_1=p_2=1/2$. Here the ferromagnetic (paramagnetic) phase is labeled as $0$ $(1)$.}
\label{fig:fig3}
\end{figure}

Our aim is to understand whether transfer learning is possible using QNN. More specifically, all our training data consists of $\kappa=0$ and we use that to predict phases at regions where $\kappa \geq 0$. This is particularly relevant since for $\kappa=0$ the model is analytically solvable, pointing out that a transition occurs at $g \approx  1$. We highlight that for $\kappa=0$ we have only two phases:  either the ferromagnetic (phase '0') or the paramagnetic (phase '1'). However, when $\kappa \geq 0$ the ANNNI Hamiltonian also leads to a third phase, the antiphase (phase '2'), not contained in the training data. In particular, for $\kappa \geq 0.5$, we are in a region where only phases '0' and '2' are present. So, the best the classifier algorithm can do is to output '0' if the phase is indeed '0' and '1' otherwise.

\begin{figure}[t!]
\begin{center}
\includegraphics*[scale=0.6]{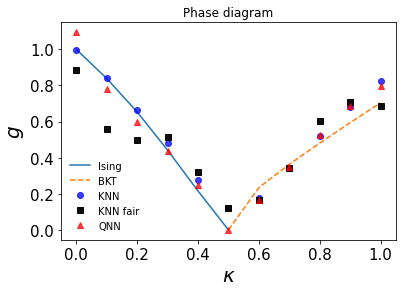}
\end{center}
\caption{
Phase diagrams produced with diverse (Q)ML algorithms when trained only with $\kappa =0$: KNN trained with raw data (blue circles); KNN trained with the same pre-processe data as the QNN - fair comparison (black squares); QNN (red triangles), and two different analytical solutions: Ising (solid blue line) and BKT (dashed orange line). 
All different methods recover the ferro/paramagnetic and paramagnetic/BKT transitions qualitatively well, although, as it is quantitatively expressed in Table \ref{tab:general_analysis}, the QNN solution yields the smallest MSE with relation to the analytical approximation, thus being an overall better solution (see main text for more details).
}
\label{fig:fig2}
\end{figure}

Actually, as discussed above, for an observation point, both the classical and quantum classifier algorithms will return a normalized probability vector $(p_0,p_1)$ where the assigned phase will correspond to the component with the largest value. As typical with such algorithms, to determine when we are facing a transition, we plot both the probability components and check when they cross, as shown in Fig.~\ref{fig:fig3}. As can be seen in Fig.~\ref{fig:fig2}, using this approach, the QNN algorithm recovers the left part ($\kappa < 0.5$) of the phase diagram, corresponding to the ferromagnetic/paramagnetic transition, very successfully. The precise prediction also holds as we approach the critical point at $\kappa=0.5$ and $\gamma=0$ at which a new antiphase appears. However, as we start to increase $\gamma$ the approximated solutions in \eqref{trans1} and \eqref{transBKT} and the QNN predictions start to differ more significantly, even though they remain qualitatively similar.

To benchmark our results we have compared the QNN solution with that obtained by the classical KNN algorithm. As can be seen in Fig.~\ref{fig:fig2}, the solution is significantly worse with the classical algorithm is fed with the same pre-processed data as the one given to the quantum algorithm. However, if the classical algorithm uses the complete data (not pre-processed) it reaches a similar success in the prediction, even though it is smaller as quantified by shown ahead. Importantly, the quantum classifier performs significantly better at the critical point $(g=0, \kappa=1/2)$.

\begin{table}[!t]
\caption{Performance (average $\ell_2$-norm with relation to the analytical approximations given by Eqs. \ref{trans1} and \ref{transBKT}) computed for the three main phases and comparing QNN with KNN (using both the pre-processed and complete training data). For the classical KNN we used $k = 7$ and the Euclidean distance. The best result is in boldface. 
}
\label{tab:general_analysis}
\setlength\tabcolsep{0pt} 
\footnotesize\centering
\smallskip 
\begin{tabular*}{\columnwidth}{@{\extracolsep{\fill}}lcc}
\hline
\diagbox[width=8em]{Technique}& average $\ell_2$-norm \\ 
 \hline
 \hline
QNN (pre-processed) & \textbf{0.0036(6)} \\
KNN (pre-processed) & 0.0164(1) \\
KNN (raw data) & 0.0043(1) \\
\end{tabular*}
\end{table}

\section{Discussion}
\label{sec:disc}


The detection of phase and phase transitions of the ANNNI model with both classical as well as quantum heuristic approaches have already been done. In Ref. \cite{canaPRB}, Canabarro et al. showed the possibility of applying both unsupervised and supervised classical techniques to achieve good results. In fact, the problem was satisfactorily well solved with unsupervised learning, reducing the use of supervised learning to a validation step. Therefore, they tried using transfer learning of diverse supervised learning algorithms trained solely on nearest-neighbor interactions exhibiting the capacity to discern a novel phase emerging upon the introduction of next-nearest-neighbor interactions. They showed that some of the learners could unveil the Ising as well as the antiphase and the paramagnetic phases. This amalgamation effectively groups the BKT and CIC transitions together, showcasing the robustness of our method. On the other hand, in Ref. \cite{monaco}, Monaco et al. used quantum convolutional neural networks by training only on marginal points of the phase diagram represented by integral models. Our approach in this paper is both innovative and complementary. We use a new and simpler quantum machine learning algorithm and apply transfer learning, we test some ideal preprocessing of the data to fit in a real quantum computer, and we train only on $\kappa =0$, testing on the remaining phase diagram to show the viability of the technique as we discuss here.

We show that with the right pre-processing of the data, the quantum nearest neighbor (QNN) algorithm proposed in \cite{schuld} allows for the classification of phases in a non-trivial Hamiltonian model. Using two-point correlation functions obtained by exact diagonalization on a small $N = 12$ spins lattice, we could reproduce the main phases of the axial next-nearest neighbor Ising (ANNNI) model. More precisely, using as training data only the ferromagnetic and paramagnetic phases, we could detect a transition for an antiphase by increasing the interaction strength of the next-nearest neighbor. This is a relevant instance of transfer learning, since using analytical data extracted from the exactly solvable transverse field Ising model, we could explore a non-integrable region of the Hamiltonian model. This makes the approach computationally cheaper as access to training labels is one of the major bottlenecks for any supervised method.


To benchmark the quality of our quantum machine model, we compared it with approximated expressions obtained by various methods. The solution provided by QNN works very well in the ferromagnetic and paramagnetic regions, offering a less precise but still qualitatively reasonable solution as we enter the antiphase. Arguably, however, to assess the quality of a quantum learning method, it is reasonable to compare its performance with that of classical learning algorithms. We performed this comparison, and the results are quite favorable to the quantum approach. Even when we feed the original data (without any pre-processing, a necessary step to reduce the number of qubits in the quantum realization) to classical classifiers, the quantum classifier remains superior, as can be seen in Fig. \ref{fig:fig2} and Table \ref{tab:general_analysis}. And performing the fairest comparison, obtained when the quantum and classical algorithms see the same pre-processed data, the accuracy of the quantum classifier is significantly higher. Importantly, these performance comparisons were done on the testing data, that is, we were really evaluating the generalization capability of the different models, which is, after all, what matters the most when one builds a data-driven model. 

This proof-of-principle (since it was obtained in a simulated/perfect quantum circuit) quantum advantage does not come in terms of algorithmic complexity, but rather in generalization and accuracy, which is of major interest in the context of machine learning. Still, one may interpret the advantage from a different point of view, namely that of sample complexity \cite{vapnik, mohri, mello}: the quantum algorithm could find the underlying pattern reflected on the dataset with much less information than its classical counterpart, and with better generalization performance. As mentioned before, we can see that as a way of reducing the dataset only to the most representative samples and better explanatory features. Although we focus on a particular Hamiltonian, we believe it leads to relevant general questions: in a statistical learning theoretical sense, how and why such a sample complexity reduction and consequent quantum advantage is achieved? Similar questions have been addressed in recent research \cite{cao_1121, caro_0621, maria_expressive}, and further research along this direction, in the particular context of QNN might lead to new insights. Another clear path is to understand how well the QNN classifier works in real NISQ devices, also considering different Hamiltonian models and increasing the number of qubits and features. In this regard, it would be interesting to consider other data formats, such as the classical shadows \cite{classical_shadows}, efficiently encoding classical data about quantum systems for machine learning purposes \cite{encoding_classical_data}.
  




In conclusion, to the best of our knowledge, this paper is the first work in which the QNN algorithm \cite{schuld} was applied to a concrete classification problem in the context of condensed matter physics. By applying this method, we could achieve a quantum model whose generalization performance was superior to its classical counterparts, whilst using much less information, which represents a quantum advantage in both contexts of generalization and sample complexity. This is the main result of this paper, which opens the way to several discussions concerning the statistical learning theoretical properties of the QNN model.


\section*{Acknowledgements}
This work was supported by the Serrapilheira Institute (Grant No. Serra-1708-15763), the Simons Foundation (Grant Number 1023171, RC), the Brazilian National Council for Scientific and Technological Development (CNPq) via the National Institute for Science and Technology on Quantum Information (INCT-IQ) and Grants No. 307172/2017-1, the Brazilian agencies MCTIC, CAPES and MEC. AC acknowledges a paid license by the Federal University of Alagoas for a sabbatical at the University of São Paulo, and partial financial support by CNPq (Grant No. $311375/2020-0$), Alagoas State Research Agency (FAPEAL) (Grant No. APQ$2022021000153$) and São Paulo Research Foundation (FAPESP) (Grant No. $2023/03562-1$).

\bibliography{refs}

\section*{Appendix}




With the aim of making the paper as self-contained as possible, in this section, we provide a brief description of classification tasks in machine learning \cite{knn_orig}, followed by a presentation of the classical KNN classifier, and the ensemble methods used in the feature selection procedure.

\subsection{Classification Problems}
\label{sec:sclassif}

In a classification problem or task, the algorithm has to assign one out of a number of discrete classes to an observation, according to a rule learned from a set of labeled (previously classified) examples. Exemplary problems include; i) diagnosing a disease given a number of symptoms; ii) credit decision based on features of the applicant; iii) classifying an email as "spam" or "ham (not-spam)", given its content; iv) recognition of a handwritten digit; v) identification of different phases in many-body systems, to cite a few.

Classification problems are addressed within the supervised learning paradigm, in which a data point $p$ is characterized by a $n$-dimensional data vector $\bm{x}^p$, whose components are $x_{i}^p$ with $i=1,\cdots, n$ (which we refer to as the \emph{features}, and may be given by binary, integer or real-valued numbers); as well as by its respective class assignment $y^p$ (which we refer to as \emph{target}). We say that the feature vector belongs to the \emph{feature space} $\mathcal{X}$, i.e., $\bm{x}^p \in \mathcal{X}$; and the target belongs to the \emph{target space} $\mathcal{Y}$. Classes, which are discrete, are often encoded by a finite number $d$ of positive integers, that is, $y^p \in \mathcal{Y} = \{0,...,d-1\}$. A particular case of interest is that of binary classification, in which case $d=2$, and $y^p \in\{0, 1\}$. Within the supervised setting, machine learning requires that $N$ samples of features and their respective targets are collected --- reason why such data points are often called \emph{observations}. Together, a set of sample observations make up the \emph{training dataset}, $\mathcal{T}=\{ (\bm{x}^{p}, y^p) \}_{p=1}^N$.

From the perspective of statistical learning theory \cite{vapnik}, a supervised learning problem may be posed as follows.
It is assumed that there is a process that relates the features and the target, i.e., that determines $y$ from $\bm{x}$. We can mathematically describe such a process with the so-called \emph{target function} $\mathcal{F} \colon \mathcal{X} \to \mathcal{Y}$.\footnote{To account for noisy targets, it is in fact necessary to introduce the \emph{target distribution} $P(y \mid \bm{x})$. In this setting, a noisy target may be given by a deterministic portion $\mathcal{F}(\bm{x}) = \mathbb{E}(y \mid \bm{x})$ as well as by the noise $y - \mathcal{F}(\bm{x})$. If there is no noise, a special case occurs for $P(y \mid \bm{x}) = 0$ everywhere, except for $y =  \mathcal{F}(\bm{x})$.} 
In principle, at least in a theory-driven approach, it could be possible to construct $\mathcal{F}(\bm{x})$ from first principles. Most often, however, the problems under consideration are too complicated for a theoretical description to be feasible. In such cases, we resort to a data-driven approach, which is commonly implemented with machine learning techniques. In this case, the so-called \emph{hypothesis function} $f_{\mathcal{H}, \bm{w}}(\bm{x})$ is introduced, which is intended to approximate the theoretical process $\mathcal{F}$. One important point is that the hypothesis is parameterized by the \emph{parameters vector} $\bm{w}$. In this setting, $\mathcal{F}$ is unknown, and our job is to approximate it with $f_{\mathcal{H}, \bm{w}}$, i.e., we want $f_{\mathcal{H}, \bm{w}} \approx \mathcal{F}$. In order to do so, we are presented with the training data $\mathcal{T}=\{ (\bm{x}^{p}, y^p) \}_{p=1}^N$, which we assume to have been generated by $\mathcal{F}$, i.e., $y^p = \mathcal{F}(\bm{x}^p)$.

An important concept is that of a \emph{hypothesis set} $\mathcal{H}$, which is the set of particular functional dependencies that we allow the hypothesis to take. That is, it is the set of all candidate hypotheses, so that $f_{\mathcal{H}, \bm{w}}(\bm{x}) \in \mathcal{H}$. Since the functional dependency among candidates on $\bm{x}$ is the same, it is clear that different candidate hypotheses differ only by the parameters vector $\bm{w}$. It is the role of the \emph{learning algorithm} to take a hypothesis set alongside with the training data, and then determine the values of the particular parameters such that on the training dataset the hypothesis is as close as possible to the real process, i.e., $f_{\mathcal{H}, \bm{w}}(\bm{x}^p) \approx \mathcal{F}(\bm{x}^p$) \footnote{Operationally, the determination of the particular $\bm{w}$ that best fits $f_{\mathcal{H}, \bm{w}}$ to $\mathcal{T}$ is achieved via the solution of an optimization problem, namely, the optimization of the \emph{cost function}, which is an error function, accounting the differences between predicted targets $f_{\mathcal{H}, \bm{w}}(\bm{x}^p)$ and actual targets $y^p$. For details, see \cite{mohri, mello}.}. At the end, a single hypothesis is picked, via the determination of the particular $\bm{w}$. Together, the hypothesis set $\mathcal{H}$ and the learning algorithm make up the so-called \emph{learning model}.

Therefore, machine learning methods aim at estimating $\mathcal{F}$ based on limited knowledge of it, reflected on $\mathcal{T}$. In other words, the learning algorithms try to capture the underlying pattern of the population, based on what is observed on the sample. Naturally, in such a scenario, one major concern is the \emph{generalization} of such an approach, i.e., how confident we are that $f_{\mathcal{H}, \bm{w}}$ as trained using data from the training set, will perform well as a stand-in for $\mathcal{F}$ on data \emph{that was not seen in training}. This question is extensively addressed via tools of statistical learning theory \cite{vapnik, mohri, mello}.

Pragmatically, solving a problem with a machine learning approach is a good idea if:
\begin{itemize}
    \item There is data available for the problem at hand - for an empirical modeling approach such as machine learning, data is absolutely essential. Without sample data to learn from, it is simply impossible to apply any machine-learning technique;
    \item A pattern exists and is reflected in the sample - after all, if there is no pattern to be learned (or, pragmatically, if the pattern is not captured in the training dataset), not even the best learning technique will be able to produce a good model;
    \item A theoretical description is not available - there will always be errors associated with a machine learning model since it is a method to empirically estimate the true process. Therefore, if a full theoretical description is available, its use would be preferable, since such estimation errors would not be present.
\end{itemize}

The aim of our work is to show that a quantum classification algorithm can perform well in the classification of phases of a non-trivial Hamiltonian, the ANNNI model, a problem that fits the requirements listed above. The training data is given by the correlations between the spins of different sites in the network. Those features are experimentally accessible and form the basis of any attempt to classify phases in many-body systems, thus fulfilling the first two criteria listed above. As for the third criterion, as discussed in Section \ref{sec:model}, there are no analytical tools for witnessing phase transitions in general, a problem that typically relies on perturbation theory, approximations, or exact diagonalization on small chains.

\subsection{K-Nearest Neighbors classifier}
\label{sec:knn}

We now present the fundamentals of the K-Nearest Neighbors (KNN) classifier \cite{knn_orig}, the classical counterpart of the QNN algorithm described in Sec. \ref{sec:QNN}.
 
\begin{figure}[t!]
\begin{center}
\includegraphics*[width=\linewidth]{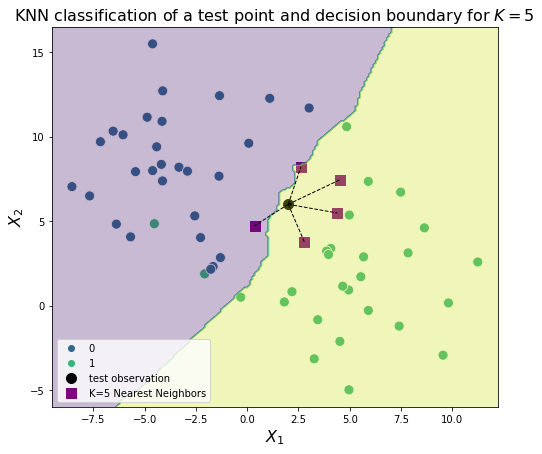}
\end{center}
\caption{\textbf{Pictorial Illustration of KNN} Considering $K=5$. Based on the distances from the 5 training points closest to the input test observation, this input would be classified as class 1 (green). The decision boundary (surface separating the 2 classes) is also depicted.}
\label{fig:knn_3}
\end{figure}

Given a training dataset $\mathcal{T}=\{ (\bm{x}^{p}, y^p) \}_{p=1}^N$ and an unclassified observation $\bm{x_{\text{in}}}$ encoding $n$ features, the goal is to assign to the input vector $\bm{x_{\text{in}}}$ a given class in $\mathcal{Y}$, using information from the training data. A possible approach for such a classification is to use some distance measure and then assign to the input vector the class whose members are closest to it.
That is the essence of the KNN classifier, with the particularity that only the $K$ nearest neighbors are considered for the classification (see Fig. \ref{fig:knn_3}), i.e., the input observation is predicted to belong to the class which is the most representative among its $K$ nearest neighbors.

In short, the KNN algorithm proceeds via the following steps: 1) set the value for $K$ (the number of neighbors to be considered); 2) For every instance $x$ (query example) in the training set: i) calculate the distance between the query example and true output, ii) add the distance and the corresponding index of the query example to an ordered collection; 3) Sort the ordered collection of distances and indices from smallest to largest (in ascending order) by the distances; 4) Pick the first $K$ entries from the sorted collection; 5) Get the labels of the selected $K$ entries; 6) Return the mode of the $K$ labels (classification tasks).

Apart from $K$, which is an obviously important hyperparameter, it is also crucial to choose an appropriate distance metric, so that the determination of neighbors' observations is meaningful. Several distance metrics are possible, each exploring a particular aspect of the dataset structure. In this work, we will consider the Hamming distance, which is used when the features are encoded as binary strings or vectors. It is defined as the number of different bits, or components, among a pair of binary strings. For instance, taking $a = 1011101$ and $b = 1001001$, we have $d_H(a,b) = 2$, since only the third and fifth components differ.

It is worth noticing that the KNN classifier is a "lazy algorithm", given that an explicit training procedure does not take place --- which also means that no explicit hypothesis from a hypothesis set will be determined. In fact, each input observation is classified at a time. This is also the case for the Quantum Nearest Neighbors algorithm.

We trained the KNN model using sklearn's implementation \textit{sklearn.neighbors.KNeighborsClassifier} \cite{scikit}. We ran a cross-validated grid search to select the best values for $K$, whilst we used the Euclidean distance for the model trained with the raw features, and hamming distance for the model trained with the pre-processed, binary data.

\subsection{Random Forest and Extra Trees} 
\label{sec:rf}

Random Forest \cite{breiman} is an ensemble method used for classification tasks, consisting of a multitude of decision trees. In the end, it outputs the most voted class of the individual decision trees, mimicking the ancient idea behind the "wisdom of the crowds". Decision tree learning consists of the construction of a decision tree from labeled training instances. The root is the training data itself. The branches are the output (Boolean output) of tests for the attributes. Finally, the leaf nodes are the estimated class. Therefore, for a given input $x$, a label $y$ is predicted after percolating from the root up to the leaf. It must be noted, however, that an individual decision tree algorithm is a weak learner since it produces high-bias errors.

To implement a combination of decision trees, we first partition the training set in $M$ smaller subsets ${B_1, B_2, ..., B_M}$, that are randomly chosen and may be repeated. If the subsets are large enough for training a specific learner, they can be aggregated to create an ensemble predictor. For classification tasks, we take a majority vote on all the predictions. This process was introduced in Ref. \cite{breiman_dt} and is called $\textit{Bagging}$, an acronym for Bootstrap Aggregation, and was shown to reduce the variance (out-of-sample error) without increasing the bias (in-sample error) \cite{highbias}.  

Within this context, the key idea of Random Forest is to perform subsets of the features. Now, the trees randomly choose a $k$ number, of $N$ total features with $k<N$. This bagging of features reduces the correlation between the various Decision Trees, contributing to better modeling. The code employed in our work was made using the package \textit{sklearn.ensemble.RandomForestClassifier} \cite{scikit}, using parameters given by \textit{max\_depth}=None, \textit{n\_estimators}=1000.

Operationally, the calculation proceeds as follows. For every tree in the ensemble, we traverse its internal nodes and calculate the impurity reduction that the respective split produces. Naturally, the impurity reduction depends on the criterion used to measure the quality of a split. For classification problems, the scikit-learn implementation supports the Gini impurity and the information gain (entropy) criteria. For this reason, even though the "\emph{impurity} reduction" term is often used, if we use the entropy criterion, we are not dealing with a proper impurity, and should actually use the term "error reduction".

The impurity reduction $\Delta_{\epsilon}$ is calculated as the difference between the impurity before the split and after the split (which is given by the averaged impurities of each one of the two branches), that is, $\Delta_{\epsilon} = \epsilon_{\text{before split}} - \epsilon_{\text{after split}}$. This impurity reduction is multiplied by the proportion of observations passing through the node, which is how the average becomes weighted. 
Remind that in each node we have a feature, therefore, the reduction is associated with a given feature. 
After this procedure, we have a given score for each feature, when we analyze a single tree. To get the final scores, we average the scores over all trees in the ensemble. Notice that the features that more commonly appear in the nodes will naturally have a higher total score, which intuitively makes sense: if a given feature appears in many nodes, and is responsible for a higher impurity decrease, it is a "better" feature, for the purpose of classification. 



As mentioned in Sec. \ref{sec:methods_qnn}, we employed a modification of Random Forest, known as Extremely Randomized Trees (Extra trees). The major difference is that, in the case of Extra trees, additional randomness is induced in the way of computing the tree splits: instead of using a fixed better-separating threshold (which is the case for traditional Random Forest), we randomly generate thresholds for each candidate feature for the split. Introduced in \cite{extra_trees}, the algorithm yields smaller variance, although the bias is often slightly increased. But, for the purpose of feature selection (which is how we employed this algorithm), it is often a better choice, since the extra randomness allows us to choose the features by their actual discriminative power, avoiding the undesired effect of a fixed threshold.

In practice, once we have an importance score for each feature, we can pick the ones that are the most important. We use the implementation from scikit-learn \cite{scikit}, and chose the $n=4$ most important features.

\subsection{Illustration of the quantum circuit}
\label{sec:qc}

\begin{turnpage}
    \begin{figure}[htb]
\includegraphics[width=\linewidth]{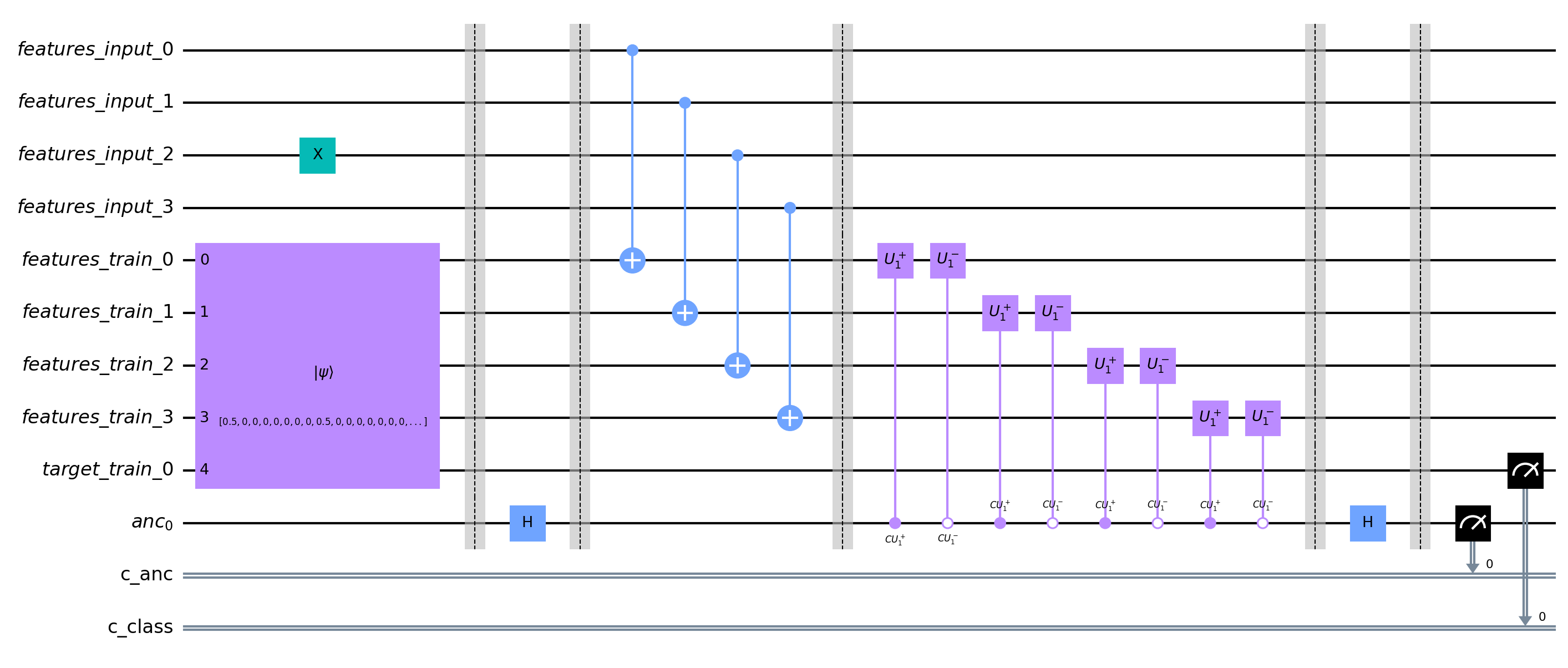}
\caption{An illustration of the QNN quantum circuit for a particular dataset of $N=4$ observations and $n=4$ features. We plot a smaller circuit here for the sake of space. The test point is given by the binary string 0010. Each barrier in the plot indicates an important step of the quantum algorithm (as described in Sec. \ref{sec:QNN}), after which the quantum state is the one given respectively by Eqs. \ref{eq:psi0}, \ref{eq:psi1}, \ref{eq:psi2}, \ref{eq:psi3} and \ref{eq:psi4_rewritten}. The measurement yields the probabilities as described in Eqs. \ref{eq:p_0} and \ref{eq:p_1}.}
\label{fig:qc}
    \end{figure}
\end{turnpage}

In Fig. \ref{fig:qc} the QNN circuit is illustrated for the following training dataset, consisting of $N=4$ observations of $n=4$ binary features:

\begin{table}[!h]
\begin{tabular}{cccc|c}
\multicolumn{4}{c|}{X} & y \\ \hline
0 & 0 & 0 & 0 & 0 \\
0 & 0 & 0 & 1 & 0 \\
1 & 1 & 1 & 0 & 1 \\
1 & 1 & 1 & 1 & 1
\end{tabular}
\end{table}

\noindent and the test observation $0010$. The superposition state of all the training datapoints (Eq. \ref{eq:train_sup}) is represented by the black-box state preparation routine given by the \textit{initialize} Qiskit instruction. In practice, this instruction is decomposed into elementary gates during the circuit transpilation.

Notice how the full circuit for this problem instance uses $2n + 2 = 10$ qubits and 2 classical registers and has a depth of 13 (which considers the state preparation routine as depth 1 - in practice, the superposition construction takes time $\mathcal{O}(Nn)$ \cite{associative_memory}). Indeed, that matches the $\mathcal{O}(n)$ depth expected for this circuit, which mainly arises from the application of the $\mathcal{U}$ operator (Eq. \ref{eq:U_op}), which demands $2n$ sequential gates.

Finally, we chose to illustrate the quantum circuit for this simpler problem instance instead of illustrating one of the actual quantum circuits built in the study, because those have 18 qubits, which would be much more difficult to visualize.

\clearpage
\clearpage
\global\pdfpageattr\expandafter{\the\pdfpageattr/Rotate 90}
\clearpage

\end{document}